\documentclass[conference,compsoc]{IEEEtran}
\ifCLASSOPTIONcompsoc
  \usepackage[nocompress]{cite}
\else
  \usepackage{cite}
\fi
\ifCLASSINFOpdf
\usepackage[pdftex]{graphicx}
\DeclareGraphicsExtensions{.pdf,.jpeg,.png}
\else
\fi
\begin{document}
\title{Towards Optimised Data Transport
and Analytics for Edge Computing}

\author{\IEEEauthorblockN{Phil Lane and Richard Hill}
\IEEEauthorblockA{Department of Computer Science\\ School of Computing and Engineering\\
University of Huddersfield\\
Huddersfield, UK.\\
Email:\tt{\{p.lane\},\{r.hill\}@hud.ac.uk}}}

%


% compsoc conferences compared to traditional conferences), use this
%Telephone: (800) 555--1212, Fax: (888) 555--1212}
\maketitle

\begin{abstract}
Industrial organisations, particularly Small and Medium-sized Enterprises (SME), face a number of challenges with regard to the adoption of Industrial Internet of Things (IIoT) technologies and methods. The scope of analytics processing that can be performed on data from IIoT-enabled industrial processes is typically limited by the compute and storage resources that are available, and any investment in additional hardware that is sufficiently flexible and scalable is difficult to justify in terms of Return On Investment (ROI). We describe a distributed model of data transport and processing that eases the take-up of IIoT, whilst also enabling a capability to securely deliver more complex analysis and future insight discovery, than would be possible with traditional network architectures.\\
\end{abstract}


% \fi
\IEEEpeerreviewmaketitle
\section{Introduction}
As costs associated with computation and storage resources continue to decrease, along with developments in the deployment of mathematical techniques for data analytics, there is considerable interest from the business community at the prospect of using operational data to increase productivity, profitability and to facilitate economic growth.

%Industrial enterprises have traditionally been at the centre of wealth generation, particularly the manufacturing sector, where raw materials are converted into finished goods.

Manufacturing industries generate vast quantities of data that can be processed using a variety of established and emerging techniques to discover new insight into how a set of operations might be modified to improve performance in the future.

Whilst the manufacturing industry contains large corporations with brand names that are recognisable to consumers, the bulk of the sector is dominated by Small to Medium sized Enterprises (SMEs), who often specialise in products, processes or markets, typically as suppliers to other businesses (B2B) as parts of complex, interconnected supply chains.

%Organisations that have already embraced technology to increase the visibility of their operations already report an ability to secure increased competitive advantage.

%The first stage of the journey towards 
Greater operations awareness can be accomplished by applying analysis techniques to streams of data that are currently being collected as a by-product of a particular operation, but as yet, the organisation has found no use for the data. This phenomenon of redundant computation and storage capacity tends to exist in larger enterprises, and is thus an expedient way of quickly developing a capability for manufacturing analytics.

In contrast, SMEs have more constrained resources, requiring compelling business cases for capital investment. An SME may find it easier to justify investment into manufacturing plant, where the ROI is predictable and tangible, rather than indirectly into equipment that offers the prospect of increasing the efficiency of existing plant and operations.  As SMEs tend to have no experience of the benefits that analytics can bring, they are reluctant to adopt the second approach.

The development of inexpensive and innovative solutions for the automation of sensing equipment is increasing awareness of what the possibilities of data analytics can do, especially since this equipment can be retrospectively fitted (relatively simply) to manufacturing plant that at present does not automatically collect or report data about its operations.

The Industrial Internet of Things (IIoT) is an emerging area that is gaining prominence as a potential route forward for industrial enterprises who wish to engage with the Industry 4.0 movement; this requires enterprises to think about their operations from a digital perspective, to harness the potential productivity benefits of collecting, analysing and sharing data across manufacturing processes, facilities, supply chains and even whole industries.

One issue that is regularly cited by industrial enterprises who are exploring the use of IIoT, is that of data security. Organisations have considerable value within their data, in the form of Intellectual Property (or `trade secrets’). 

Reticence towards the adoption of inexpensive utility solutions such as cloud computing and Software as a Service (SaaS) is typical, as there is the perception that new security vulnerabilities will be introduced into the fabric of the business, whether it is the stealing of intellectual property, or the potentially devastating effects of service disruption from third parties. Indeed, Gartner in their IIoT Magic Quadrant report \cite{Gartner2018} recognise the importance of industrial analytics solutions that do not rely on cloud solutions.

Thus, businesses need an inexpensive, convenient, flexible and scalable adoption route for IIoT analytics.

To do this requires an architecture that is able to optimise both data transport and processing needs, whilst maximising the secure utilisation of constrained physical resources.
\section{Development of the proposed system architecture}
%
%A typical small/medium manufacturing company will often be dated in terms of its uptake of technology.  Machinery will be old, and will not have sophisticated sensing and analytic capabilities built in.  As capital investment is hard to justify in these organisations, the only viable option is to retrofit sensing and analytic functionality.  The key question is how to best do that.

One option for retro-fitting sensing and analytics is to contract one of the major vendors to provide an out-of-the-box solution. The high cost of such solutions makes them very unattractive to most SMEs.
%This reluctance stems from the vlaue of process IP that these comanies hold, and the fear of the owners that a could solution could expopse them to potentai loss of this IP. 

\begin{figure}[b]
  \includegraphics[width=\linewidth]{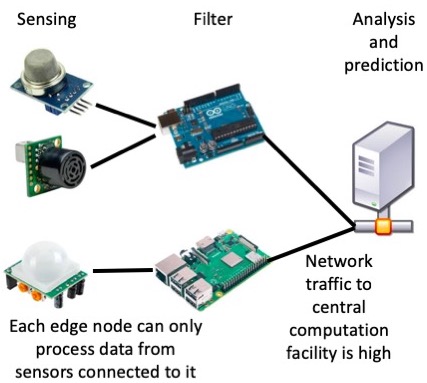}
  \caption{Edge architecture for IIoT.}
  \label{fig:edge}
\end{figure}

Given the above, there is growing interest in the use of low-cost, commercial-off-the-shelf (COTS) based solutions to enable both sensing and the processing required to implement analytics.  Typically, COTS-based solutions would involve the attachment of several sensors to, say, a milling machine, and the use of some local processing to filter and condition the sensor data prior to transmission to a networked PC where analysis and visualisation could be carried out in, for example, a spreadsheet application.  See Figure \ref{fig:edge}.  

While this approach is perfectly reasonable for small scale deployments, it does present the following problems:
\begin{itemize}
\item Lack of scalability - as more sensors are added, the physical complexity/difficulty of providing the connectivity to the PC increases with a need to add network switches, etc.
\item The joint analysis of the data from multiple sensors at edge devices an be severely limited as the data from adjacent sensors could well be attached to different edge devices; the data can therefore only be integrated `vertically' and analysed at the central PC, rather than `horizontally' across multiple edge nodes.
\item There is a real potential that the processing available at the edge devices is very poorly utilised.  Some of these devices will essentially just be passing on the data from the sensors to the central PC resource with perhaps some signal conditioning and filtering being applied.
\end{itemize}

Edge devices can be very low-cost and include, for example, devices such as the expanding range of Arduino micro-controllers or Raspberry Pi single board computers.

The solution that we are advocating here overcomes all of these difficulties and limitations.  We propose a distributed solution based on COTS components that would self-organise to enable the extraction of the maximum performance possible from the components.

\begin{figure}[t]
  \includegraphics[width=\linewidth]{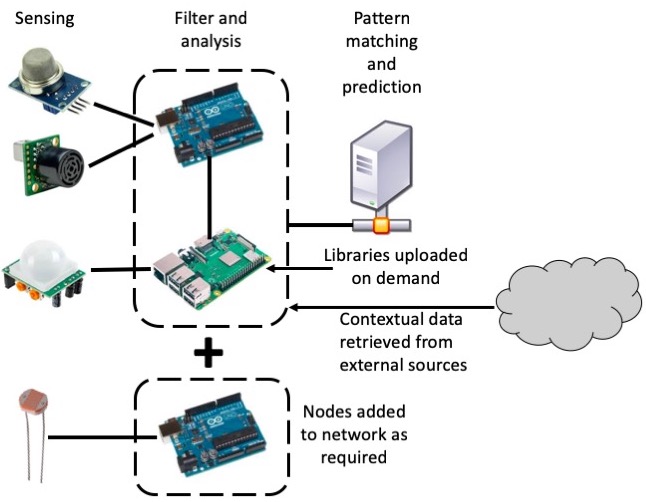}
  \caption{Overall architecture of proposed system.}
  \label{fig:arch_overall}
\end{figure}

An example deployment is shown in figure \ref{fig:arch_overall} which uses Arduino micro-controllers at the edge of the network to consolidate and condition the data from the sensors.  A more capable device such as a Raspberry Pi would enable analysis to be performed on the sensor data, and the display of the results of such analysis on a dashboard near the machine.  Connectivity between the edge device and the Pi could be easily implemented using e.g. Wi-Fi.

As the company starts to understand and realise the value that can be delivered by the analysis, more machines could be fitted with sensors with their own edge Arduino device, and when necessary more Pi units could be added also.  A central PC would be added when desired that could be used to run more computationally intensive predictive modelling, or for the generation of analytic models that could be downloaded to the Pis and Arduinos.

Compared to the traditional hierarchical approach, the approach proposed here has the benefit of easily enabling analysis to be performed on sensed data from multiple machines at a local level, without having to transport data to a central analysis platform.
\section{Opportunities offered by the adoption of the proposed approach}
%
%SMEs don’t trust the cloud with their process data, which is their intellectual property.
%Greater awareness of the usefulness of analytics for insight means that more businesses want to exploit this.

More pervasive technology in the form of embedded, networked devices means that there is often a significant volume of low-cost, redundant resources available in a network.

Much of the computation required for analytics can be completed on the low-cost COTS hardware advocated here, but deeper insight can often be found when a) data sources are combined; b) more iterations of processing are performed on a data set; c) the results of the processing are visualised and re-queried on demand.

Traditionally, research has concentrated on the separation of data transport from the processes required to process the data. The emerging requirement to process data locally at the edge of a network, on heterogeneous, constrained hardware that is plentiful, indicates that we need to re-consider how data is transported and processed within IIoT networks in order to optimally exploit the resources available.

The prospect of harvesting redundant compute cycles is not new in itself, though such scenarios tend to assume that standard hardware platform and networking protocols are trusted components of a system.

We foresee a need to cater for a number of situations that are relevant to an organisation's adoption of IIoT. First, there is a need for nodes with processing capability to be able to discover and be knowledgeable of the capabilities of useful neighbour nodes. In a dynamic edge network, it is not necessarily feasible to assume the every node is aware of every other node in the network. 

Second, there is a trade-off (or series of trade-offs) in terms of whether a data packet is processed locally or transported to another node with, perhaps, more advanced computational capability. 

Third, there is a potential need to assess the actual needs of a computation job, before allocating it to a set of nodes with the appropriate capabilities, prior to governing the re-construction of the results of the processing.

Finally, there is a need to be able to upgrade the capability of a particular node by, for example, delivering library functions that implement particular mathematical operations such as an FFT.

Thus, there is a requirement to perform a number of optimisations that can a) help a node make a value-based judgement in relation to its own actions; b) provide the information to a node that wishes to delegate activities to other nodes; and c) deliver system-wide assessments that assist the balancing of workload across nodes with the requisite capabilities.

We envisage that the network would exploit a capability whereby a node with resource capacity, who lacks the software capability to perform a particular computation, would be able to `upgrade’ based on a trigger from another cooperative node.

\section{Review of literature}

Dutta et al\cite{dutta2012} provide a general review of Wireless Sensor Network (WSN) literature. Network protocols are designed to operate within severely constrained sensor node devices. Such protocols take into account energy consumption, computational and storage capabilities and communications requirements. Key behaviours are the ability to discover other nodes to form networks and the successful transportation of data. There is no explicit commentary in regard to the processing of data as a consideration.

Devanaboyina et al\cite{singhal2015} describe a parallel distributed architecture for WSN, where the sensor nodes have storage capacity only for the most recent sensed value.

The specific challenge of operating WSN in environments where it is not feasible to transport sensor data to remote data fusion centres is explored by Binder et al\cite{binder2015}. The model proposed enables observed data to be classified in real time by communicating with neighbouring network nodes, as opposed to sending each value to a master computational node.

The configuration of WSN is often subject to rapid change as nodes join and leave the network.\cite{chen2008} Unlike a static network graph, the ability to optimise at both locations and global levels is particularly challenging. Whereas graph scalability can be enabled through the use of P2P routing tables or more traditional network hierarchies, dynamic WSN must employ alternative approaches in order to provide satisfactory QoS.\cite{estrin1999,chen2006,xiao2005}

Traditional sensor networks are arranged in a star topology, whereby sensed data is stored and processed in a central server (or cloud).\cite{johnston2018} This simplifies data analysis, at the expense of transporting large amounts of data around the network. In contrast to this, Edge Computing places computational resource at the network extremities, attached to the sensors, with the benefit of reducing bandwidth consumption, as well as reducing the latency of processing local data. However, this distribution of processing does increase complexity somewhat, by requiring coordination and management controls that can govern multiple streams of data in a timely fashion.

Harth et al\cite{harth2017a} propose a model whereby data collected through edge architectures is processed within a local context, offering timely analytics of data streams. This extends the impact of edge computing by pro-actively pushing intelligent data processing to the network extremities. This is particularly suited to the requirements of time-series data analytics, where the data streams are high velocity and opportunities to gain valuable insight are temporal.

One challenge of maintaining quantities of heterogeneous hardware devices is that of software updates, to provide security, performance or capability upgrades. Herry et al \cite{herry2018} cite the scenario of intermittent network connectivity as a challenge for device maintenance, and propose a Peer-to-Peer model to eliminate the requirement of a management node.

For the industrial scenario we describe, flexibility is a fundamental characteristic, especially since there is a need for data processing tasks to be shared across a number of nodes. As described earlier, nodes will require software re-configuration from time to time, in order to be able to maximise QoS.
\section{Some observations on the problem domain}
Traditionally, data communications research has concentrated on ``further and faster'', but in the emerging application domain of IIoT, a different philosophy around communications is required.  On example of this new thinking is Multi-access Edge Computing which is a Telecom industry approach for the delivery of edge capability.

%\\sensor data production is increasing, but there is a constrained amount of bandwidth available to transport. There are also constraints in terms of power, especially for battery/solar powered devices.

%Story is: it looks feasible. And also, by doing it, it offers up the possibility to access much more computation within private domains. Companies don't trust cloud and don't want to invest in raw compute power. If they can tap into the potentially vast resource that is already available, with the benefit of that resource being scaled up as new devices are added. This saves power as well. There may exist cheaper cycles within the existing hardware (FPGA etc).

%Also opportunity is to dynamically adapt the configuration of hardware through the operation of the network, by intelligently querying the devices during execution. This is of interest to organisations who are restricting their growth potential by being reticent wrt cloud.

However, we posit that it is necessary to consider communications and data analysis together for a number of reasons:
\begin{itemize}
\item Sensor hardware is becoming more capable in terms of computation and storage.
\item There is more interest in performing analysis close to the source; processing data at the edge of the network.
\item There is a desire from IIoT users to process data within the organisation, without requiring external cloud utilities.
\item The transportation of data is expensive - wireless transmission requires more energy than wired connections - and larger data payloads require more power to transmit.
\end{itemize}
Raw data from sensors is of less value to a business than the data after it has been processed and analysed within a relevant context for the business. Sensor clouds take in data and produce intelligence. This intelligence is much more informative about the processes of a manufacturing business, and will contain comprehensible representations of manufacturing processes. These processes are often the key to an organisation’s intellectual property, and as such is something that is not to be trusted outside of the organisation.

Some organisations might be willing to embrace encrypted analytics - performed on high powered remote cloud resources - but but this requires some enterprising thinking and a more liberal attitude towards data processing.

Another aspect of trust is that it is going to be challenging to manage relationships between myriad IIoT devices. Whilst it is logical that devices within the physical bounds of a company will be identifiable and `known’, the nature of IIoT is such that devices external to the organisation can inter-operate with existing networks to enable new capabilities.

For instance, mobile sensors on the vehicles of supply chain apparatus will benefit from being able to integrate, if only temporarily, with the networks of its customers. 

As new business models emerge, a significant enabling capability of such networks is agility; this specific characteristic is something that is particularly attractive about the emerging IIoT.
\begin{figure*}[t]
  \includegraphics[width=\textwidth]{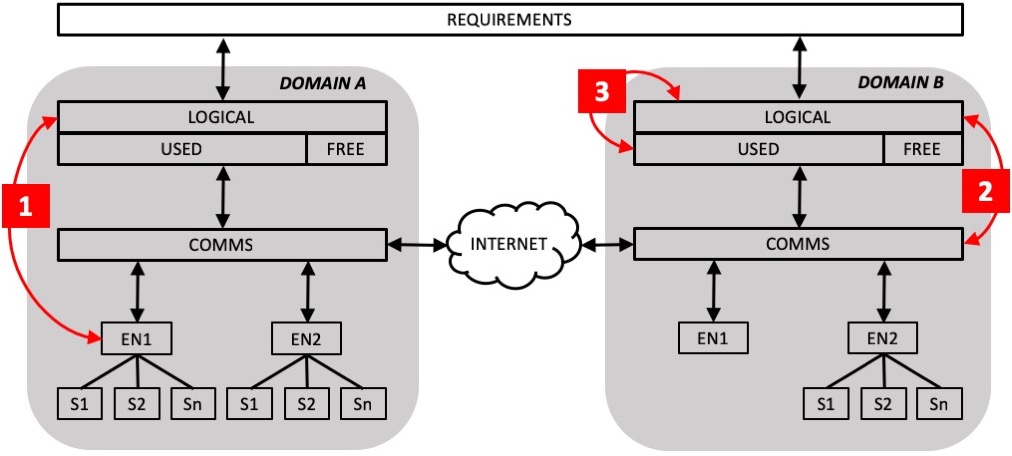}
  \caption{Model layers.}
  \label{fig:layers}
\end{figure*}
Industrial Edge network domains contain valuable intellectual property, within which lies insight for greater potential competitive advantage. Organisations need new ways of utilising the edge resources to discover process insight, in a secure and timely fashion. We are proposing a model whereby data transportation and data processing are optimised collectively, within a dynamic, heterogeneous hardware environment.

The benefits of this proposed approach include:

\begin{itemize}
\item More computational power is extracted from the hardware that is already there for a given amount of energy; this relaxes the need to purchase more hardware, but when more hardware is acquired, it is better utilised.

\item Sensitive intellectual property - the `insight' contained within sensor data that has been processed and synthesised with other data sources is analysed and retained within a security realm. This is important for organisations who do not trust other security realms, or the communication channels that exist between realms.

\item There is a more agile approach to analysis, that can cope with dynamic network architectures. This results in being able to exploit the resources of trusted mobile devices, but also over time it is feasible that data processing capabilities will be enhanced or re-configured to suit the business needs of a network at that time.

\item There is a reduction in the transport of voluminous and high speed raw sensor data across the network, reducing demands made upon shared repositories that are either on or off the premises. Only processed, insightful data is retained.
\end{itemize}
\section{Proposed overall system architecture}
Our proposed architecture consists of homogeneous nodes with a very diverse collection of processing capability (MIPS/FLOPS), memory, and an ability to run libraries and libraries already installed. The tasks to be performed are varied and not pre-definable so there is therefore a need for flexibility w.r.t. computational capability.

There exists a very wide range of computational requirements ranging from moving average calculations to complex machine learning algorithms. 
%E.g. additional capability to perform FFT on data could be one augmentation to an existing network, where the spare capacity can be utilised, after adding specific software capabilities to the network devices.\\
A fundamental assumption is that the overall architecture is not governed by one policy or set of pre-defined protocols. The operation of the network will dynamically adapt to the resource that is available, taking account of mobile (trusted) nodes that may enter or leave the network at any time. Organisations may wish to decide as to whether the data leaves the premises at all, or if it does, how much of it is released.

Figure \ref{fig:layers} illustrates the conceptual architecture of the proposed system.  The system comprises of a number of domains.  Requirements (e.g. the collation of a data set and the application. of a specific machine learning algorithm to the data) are communicated to a logical control layer.

The control layer interacts over the communications infrastructure with the edge processors. The logical control layer keeps a record of which edge devices have free capacity (i.e. spare compute cycles that can be used) and the capability of those devices (i.e. what libraries are already available and which could be installed if required).

The system at a single organisation can be distributed geographically and could potentially use the Internet to interconnect to other systems. External data can also be brought into the system to enable the extraction of more value from the internal data.
\section{Proposed node internal architecture and functional requirements}
\begin{figure}[b!]
  \includegraphics[width=275pt]{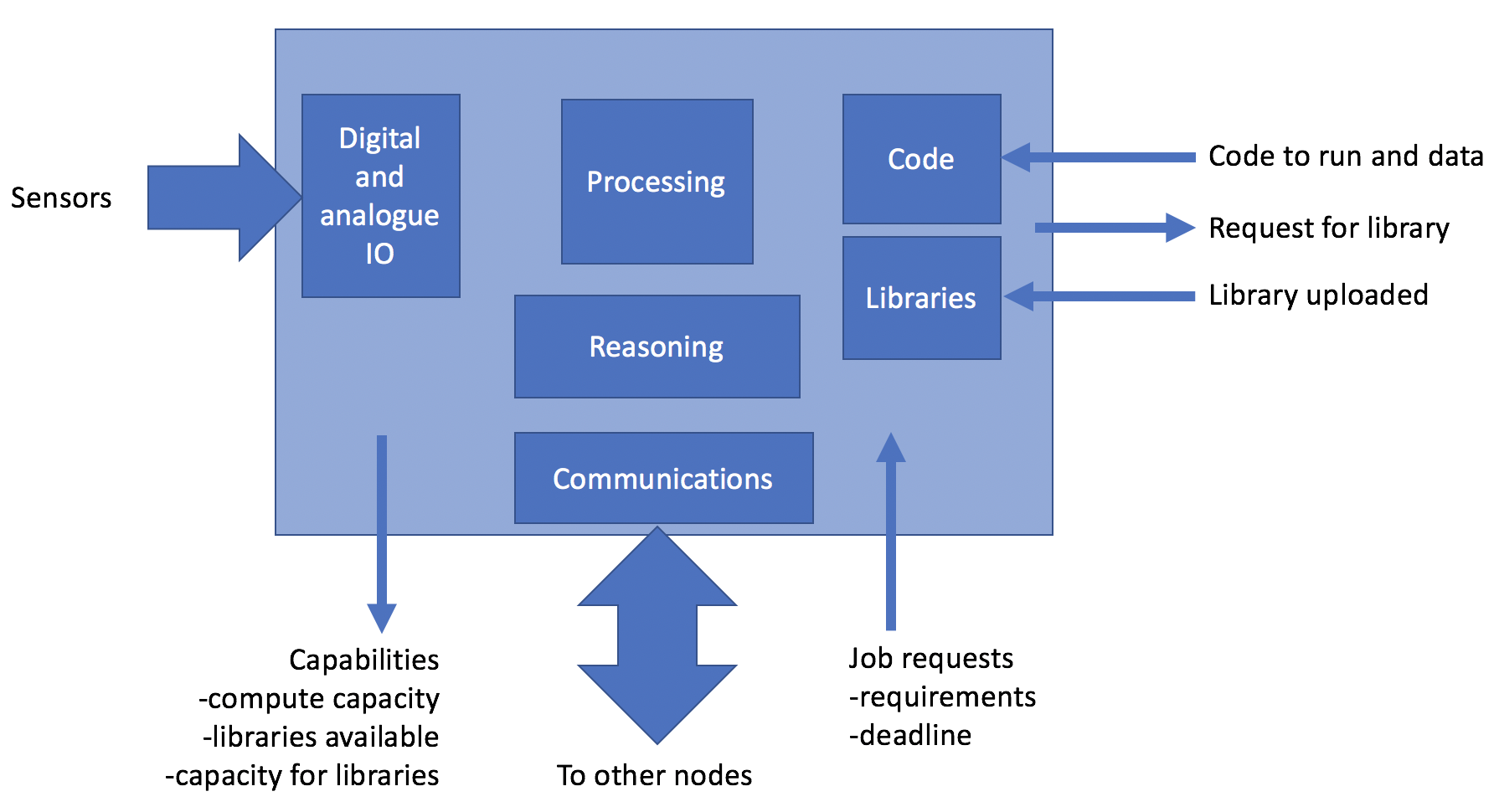}
  \caption{Proposed internal architecture of an edge node.}
  \label{fig:node}
\end{figure}
The proposed internal structure of a node is shown in Figure \ref{fig:node}. A node comprises of the following elements and functionality:
\begin{itemize}
\item Interface circuitry to connect to the sensors of relevance to that node.
\item Processing capability.
\item Reasoning to support the broadcast of capabilities, to request new libraries, to accept job requests, and to coordinate communications with other nodes and data sources.
\item Communications capability to connect to other nodes and external data sources.
\item Executable code.
\item Libraries to support the required computation.
\end{itemize}
\section{Performance optimisation}
We propose that there needs to be three optimisation loops that have to executed to ensure that the system delivers optimum performance. These are shown by the numbered loops on figure \ref{fig:layers}.  The three loops are:
\begin{enumerate}
\item \emph{Logical and processing} - partition tasks between processors to keep maximum capability free. It is important to draw a distinction between capacity (spare compute cycles) and capability (libraries installed and libraries that could be installed).
\item \emph{Logical and communications} - Knowing the task partitioning there is a need to optimise the communications to minimise energy and/or latency. Furthermore, there is interdependence between task partitioning, communications delay and power, so loop 1 needs to be aware of the trade-off; in this case a sub-optimal solution to loop 1 might be closer to optimum overall. 
\item \emph{Logical} - For a single domain this needs awareness of \emph{Tasks} and \emph{Capabilities} as it needs to ``orchestrate" both loops 1 and 2. This loop becomes more interesting if we have multiple domains (security realms) without inherent trust between them, but this is the subject of future work.
\end{enumerate}
The overall objective here is to fill the capacity with tasks, but being cognizant of keeping ``specialist" capabilities free e.g. a node that can perform FFTs should be saved for when such functionality is needed. We are therefore not proposing an optimisation to keep capacity free.

There will be a need for processing tasks to be self-describing so that the receiving node can a) decide whether it has the capability to perform the data processing request, and b) make an assessment as to the effects of the processing workload on the computational capacity that is available.

%\section{Optimization of Communications Approach}
%Two factors: latency and power
%Power linked to distance for wireless, not for wired
%Power linked to data volume - placement of computation a factor, as is compression
%Latency linked to hop cont, and whether need to send code along with data, or just data.  Physical distance is also an issue, but maybe less so within a domain unless we think of a domain as being a single intranet that could be physically disjoint/dispersed.

\section{Key issues and benefits of the proposed approach}
The requirement for data-driven decision making is increasing. This is often supported by data visualisation that is closer to the object, product or service that is being monitored.
The scope of operations sensing, and the subsequent volume of data produced is increasing.

A perceived threat to data (intellectual property) security results in a general reluctance to employ remote storage services such as cloud computing.

As data and visualisation activities become more commonplace, there is an associated increase in a business's capability to comprehend and search for new insight. As industrial data sensing capabilities develop, the introduction of new business models that exploit such technology leads to new levels of complexity that existing network architectures are challenged by.

The investment in traditional hierarchical network infrastructure is difficult to justify for an enterprise that is exploring the potential of analytics, to the point where a business case cannot be made. In contrast, this proposal describes an approach whereby capability can be incrementally added from a small starting point, presenting a low barrier to entry into the area.

The workload demands placed upon a sensor cloud/analytics system varies both in terms of a) the mix of physical activities that are being processed, as well as b) the mix of analytics activities that may need to be undertaken for a given investigation. It is likely that this computational load will be greater in the future as new applications and capabilities are introduced.

A system that can tolerate dynamic work loads must be scalable and be able to elastically provision capability on demand.
Our proposal provides the opportunity processing capability to be serviced horizontally by other nodes in the network, rather than vertically from designated computational nodes in a network hierarchy.

The adoption of such technologies will have an inevitable impact upon the human capital of an organisation, along with the associated challenges of change management. This proposal enables the capabilities to be scaled, and learned at a pace whereby the users are educated as to the tangible benefits of the technology.
\section{Conclusion}
In this paper we have proposed a novel approach to the optimum delivery of edge-based analytics that is tailored to be of specific relevance to SMEs and their particular challenges.  We now propose to build and characterise a demonstration system to prove the benefits and to explore the performance potential and trade-offs.

\end{document}